# Interstellar Photovoltaics for Exploring Alien Solar Systems

Prof. George F. Smoot
Donostia International Physics Center, Basque Country, San Sebastian, Spain
Physics Department and LBNL University of California at Berkeley *emeritus*

**ABSTRACT:** Explore alien solar systems via local star power using interstellar photovoltaics, tailored for the particular target star for maximum power and low mass. We consider tailored organic thin-film photovoltaics. Key for sensing, sending more data back and powering A.I. to send back observational summaries and interesting events and observations. This plus other technology developments are necessary for exploring Alien Solar Systems in the not too distant future.

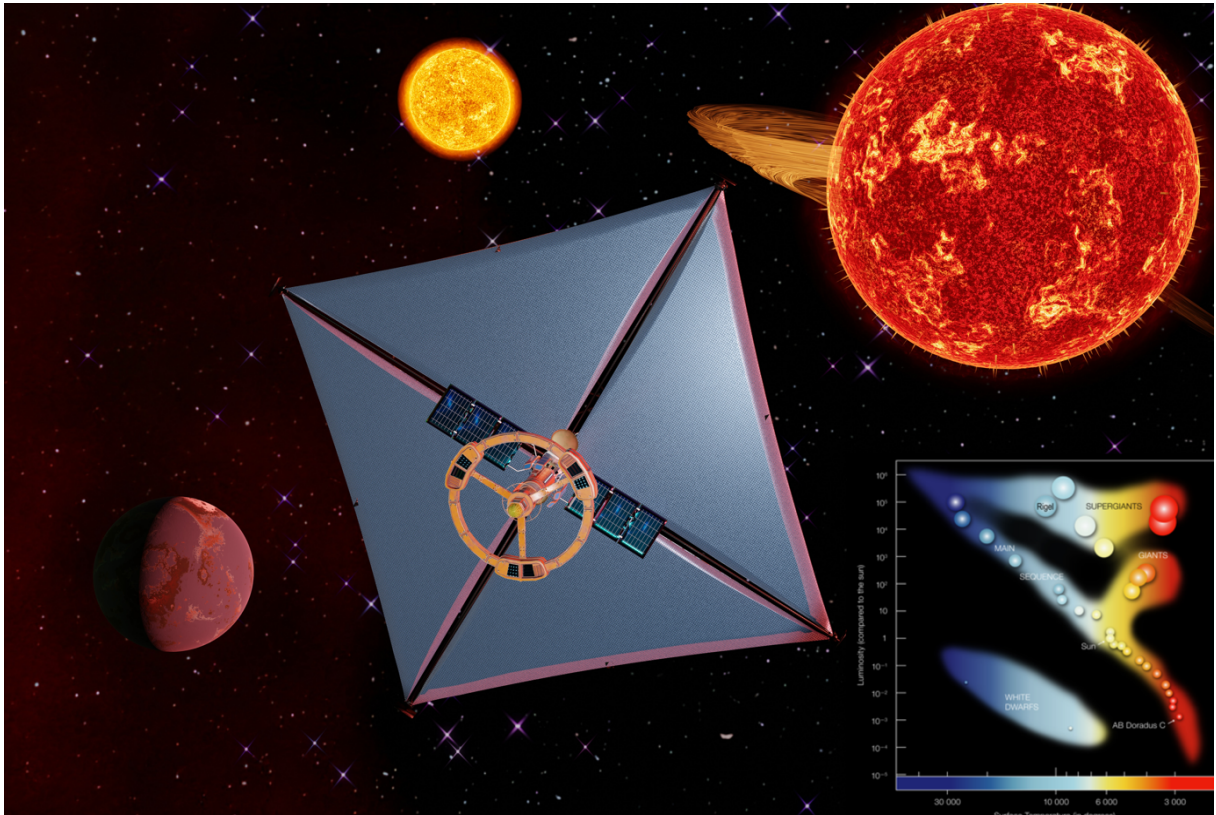

Light Sail spacecraft with organic photovoltaics and sample stars of known alien solar systems and inset of brightness versus colour temperature of all stars. Figure credit Hryhorii Parkhomenko.

**Introduction** Interstellar spacecraft need as much power as possible to make observations including imaging and data processing. The rate and also amount of data that can be transmitted back is directly related to the power available. Usually, there will be many more images taken than can be sent back due to the great distance and so more power collected by efficient photovoltaics will mean more data can be transmitted back to Earth. Images can be compressed and if sufficiently powerful, an A.I. system can help compress, and also summarize and decide which images or observations are most valuable and interesting to ensure they are transmitted. With a round trip signal time (NASA's Deep Space Network uses round trip light time as their preferred wording.) of order a decade or more, the A.I. system must decide that along with any additional manoeuvres and observations that should be undertaken.

Optimizing the efficiency, reducing the weight and overall requirements maximizes the mission. Organic photovoltaics (OPVs) are prime candidates[1] as their photoactive layers are nanometer-thin films that can be deposited on thin plastic substrates and on curved surfaces.

Further, OPVs exhibit the required mechanical flexibility and there is evidence for self-healing from radiation damage.[2]

Our work[1] addresses the possibility of harvesting photon energy from different types of stars, including Proxima Centauri. An exciting and the nearest target for exploration is the planet *Proxima Centauri b*. A member of the Alpha Centauri star system located at a distance of 4.2 light years from Earth. The planet orbits the red dwarf (surface temperature: 3,042 K) star *Proxima Centauri*, the closest star to our planet. *Proxima Centauri b* is in its habitable zone, at a distance of about 7.5 million km orbiting with a period of 11.2 days. Its nearest neighbour star Alpha Centauri is like the Sun but it is important to remember that our galaxy is also home to many hot blue stars. See the insert Hertzsprung-Russel Diagram that shows the brightness and temperature colour versus effective surface temperature of stars for examples.

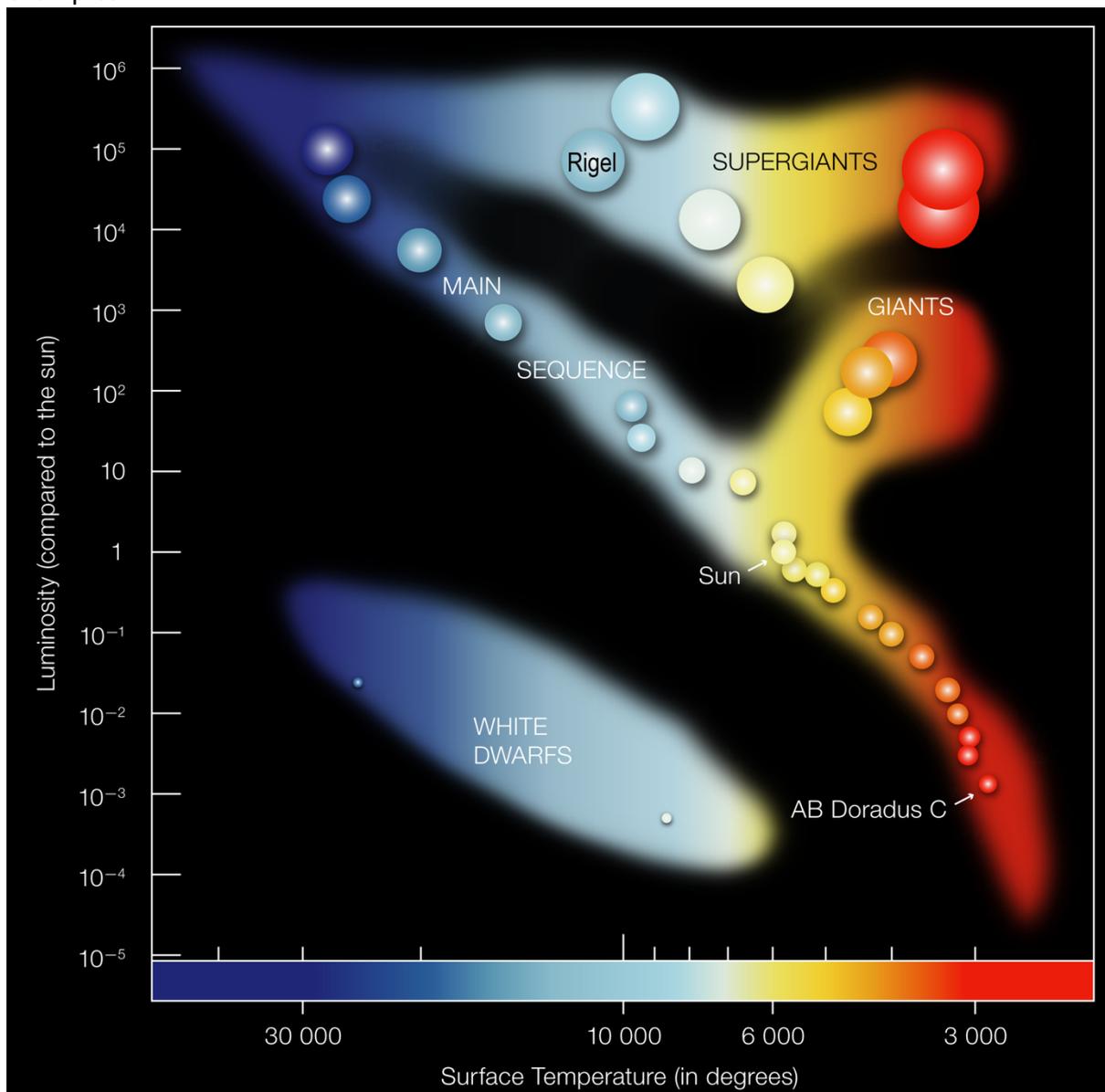

Caption: The Hertzsprung-Russel Diagram from Wikipedia commons

The spectral distribution of stars spans a wide range depending on their temperature and the Hertzsprung-Russel Diagram shows the intrinsic luminosity goes from $10^{-3}$ of solar for red dwarf stars to $10^5$ solar luminosities for the hottest blue stars. This range of $10^8$ in intrinsic power and the variation of over a factor of ten in peak flux frequency or wavelength would seem to be an insurmountable issue. However, if we are first interested in planets in the "habitable zone", where water is a liquid, then the planet's distance from the system's star puts them in a location with ambient flux within a factor of four of each other or less for the most favourable candidates.

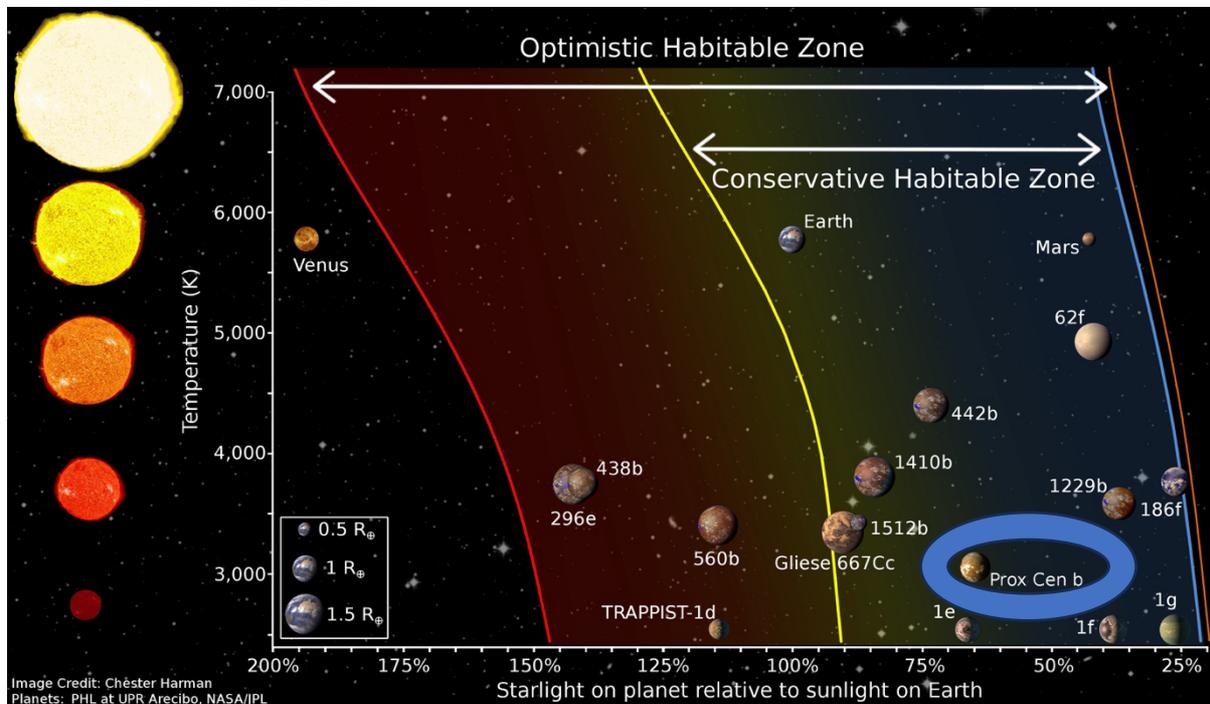

Figure showing sample known planets and potential habitable zones and their relative light power flux from their stars.

We are then left with the issue of matching the photocell material with the peak of the stellar spectrum where most of the power is conveyed.

An optimal bandgap of > 12 eV for the hottest O5V star type leads to 47% Shockley-Queisser photo conversion efficiency (SQ PCE), whereas a narrower optimal bandgap of 0.7 eV leads to 23% SQ PCE for the coldest red dwarf M0, M5.5Ve, and M8V type stars. This energy range matches the frequency range in peak flux, which is proportional to temperature ($E = h\nu = hc/\lambda \; \alpha \; kT$).

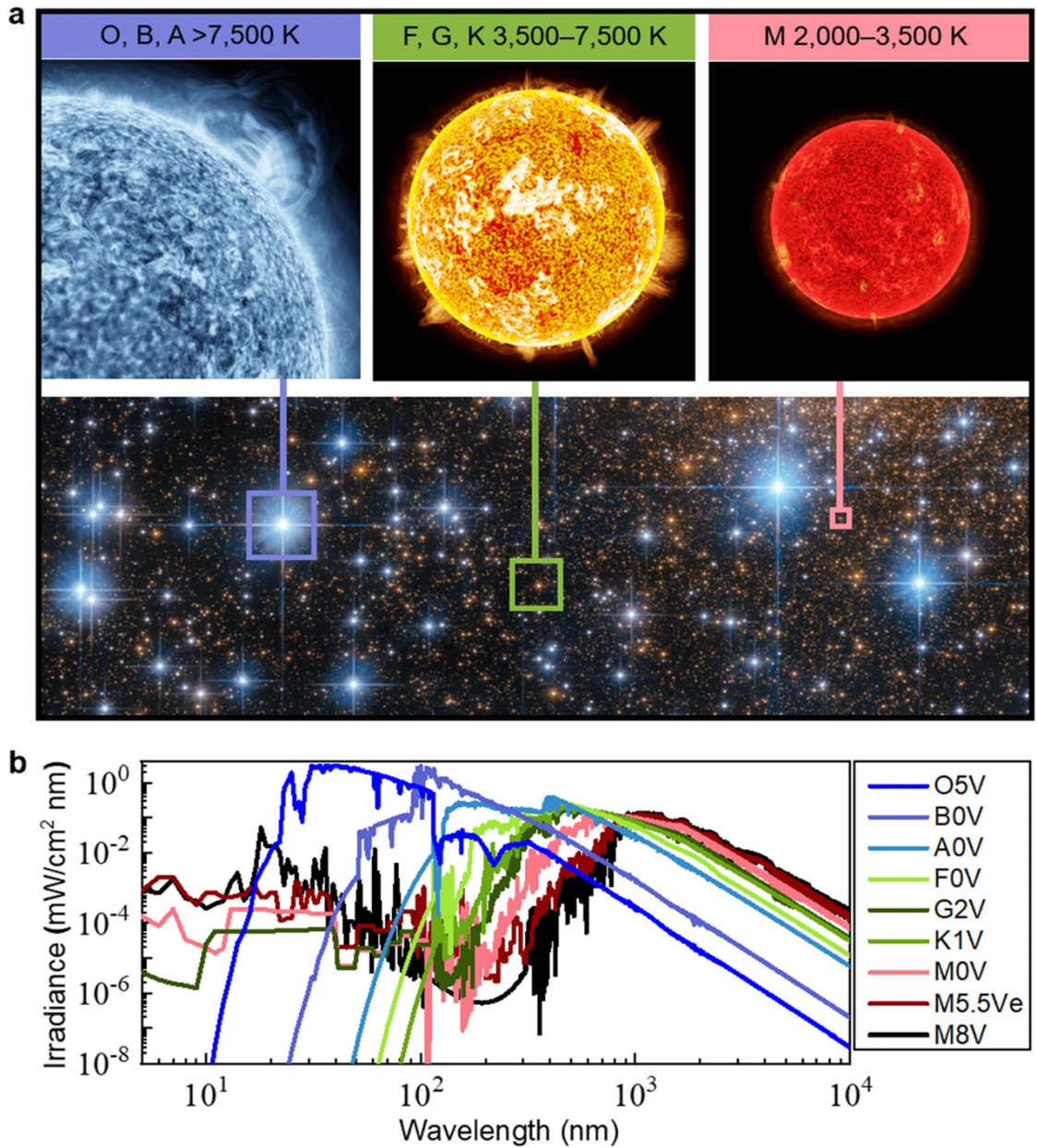

**Figure** Stellar types and their spectra. **a** Visually representative examples of O, B, A, F, G, K and M stellar types (hot to cold) within our galaxy (source: NASA54). **b** Spectral distributions of stars ranging from O5V to M8V.

We[1] considered in detail the application of a wide band gap and a narrow band gap OPV system. The first case near the Sun (G2V type star) and the other case near the red dwarf star Proxima Centauri (M5.5Ve type star.

The wider band gap system shows a theoretical PCE of 18.2% in G2V illumination that drops sharply to 0.9% under M5.5Ve illumination due to a poor spectral overlap of the absorption and the M5.5Ve spectrum. In contrast, the more extensive spectral overlap found for the PM2:COTIC-4F narrow band gap system for both the G2V and the M5.5Ve spectra leads to theoretical PCEs of 22.6% and 12.6%, respectively. Our results demonstrate the need for narrow band gap systems for interstellar OPV applications near Proxima Centauri, or, more generally, the need to consider the spectral irradiance and the material properties of OPVs for interstellar applications.

The *Breakthrough Starshot Initiative*[3] considers the aim to send a thousand centimeter-sized, gram-scale probes mounted on laser-driven 'light sails' spacecraft to the Alpha Centauri system to observe and send data back to Earth. Light sails use light propulsion from very powerful earth-based lasers via the momentum of the light or its radiation pressure. The sails' are still in development but the acceleration is limited by the ratio of laser power flux to mass per unit area. Thus, the sails must to be lightweight in order to be accelerated by lasers to a speed of ~ 0.2c, which would allow the probes to reach the Alpha Centauri within two decades of launch. Such acceleration is only possible when combining powerful gigawatt lasers with a very lightweight spacecraft of a few grams. Currently, some scientists and engineers are exploring the technical feasibility of the project, and addressing thermal management and the light sail design. The Diffractive Solar Sailing project was selected for Phase III study under the NASA Innovative Advanced Concepts (NIAC) program. NASA would like to launch such a mission on the time scale of one to two decades. In December 2017, NASA released a mission concept involving the launch, in 2069, of an interstellar probe to search for signs of life on planets orbiting stars in and around the Alpha Centauri system.[4,5]

Given the serious attention and funding being given to possible upcoming interstellar light sail missions, it is time to develop technologies that might actually be needed and utilized in addition to the high-powered laser drive. The light sail itself is needed to be developed to include the features such as the *interstellar photovoltaics* described here and also items such as making the sail provide power storage/surge by making it a supercapacitor[1] and also the system to send back the data to Earth via a laser communication system. Additional detectors such as a modernized crystal radio set[2] are highly desirable.

**Brief Summary of Breakthrough Starshot and Issues**
The big idea behind the Breakthrough Starshot project relies on advances in laser technology. This is conceivable because very high laser power output may be achieved in the relatively near future, along with improved collimation, while the cost of high-powered lasers has dropped alongside those developments.[3] As a result, it is possible to envision this ideal scenario:
- An array of high-powered lasers is constructed in space.
- A series of nanotechnology-based spacecraft are constructed, and each attached to a thin, light, highly reflective but sturdy "light sail."
- The total mass of the spacecraft and the sail, combined, comes in around a gram per square meter of sail area.

---

[1]Supercapacitors are more efficient than batteries, especially under full load conditions. They can achieve round-trip efficiency of more than 98% while lithium-ion batteries' efficiencies are less than 90%. For example, graphene has a theoretical specific surface area of 2630 $m^2$/g which can theoretically lead to a capacitance of 550 F/g. Capacitors can provide a big burst of power for data transmission or other tasks.

[2]A **crystal set**, is a simple radio receiver, popular in the early days of radio. It uses only the power of the received radio signal detected to produce sound, needing no external power. Here instead we would like to be able to process the detected radio signals to look for AM and FM modulation and signs of radio communications transmissions and have the A.I. system "listen" to the received signals. Again this requires power to be able to process the crystal (now a diode in a tuned circuit) output, perhaps sweep tune the circuit and to have the cpu and AI do deep signal processing.

[3] At the time I was working a conference in Paris with Gerard Mourou who convinced me that with advances in lasers and especially lasing fibers that could be spaced in an array for powering and then triggered to lase in synchrony with a trigger laser and tree build up would work for creating a very powerful laser beam that might be powerful enough for accelerating a light sail. It is still very expensive to think about doing in space and so the first generation proposed is for a ground-based array shown in the figure. Gerard was often asking me to participate in worthy events and conferences. When Gerard Mourou and his former student Donna Strickland were awarded the Nobel Prize in Physics in 2018, my first thoughts were well deserved congratulations and the second was that Gerard would soon learn that there were many people like he was who will want to schedule his time and efforts and then he might understand why I did not participate in all his events.

- The laser array then fires at one nanocraft at a time, accelerating it in one direction — towards its ultimate interstellar destination — to as great a speed as possible for as long as possible.
- After a journey across the interstellar medium, it arrives at its destination, where it gathers information, takes data, and transmits it back across the same interstellar distance, all the way back to Earth.

The concept stems from an announcement in April 2016, by Russian high-tech billionaire Yuri Milner. He founded a new and ambitious initiative called *Breakthrough Starshot*, with the intention to pour $100 million into proof-of-concept studies for an entirely new technology for star travel. It aimed for a speed of 20% of the speed of light, with the goal of reaching the Alpha Centauri system – and, presumably, its newly discovered planets Proxima b and Proxima c – within 20 years.

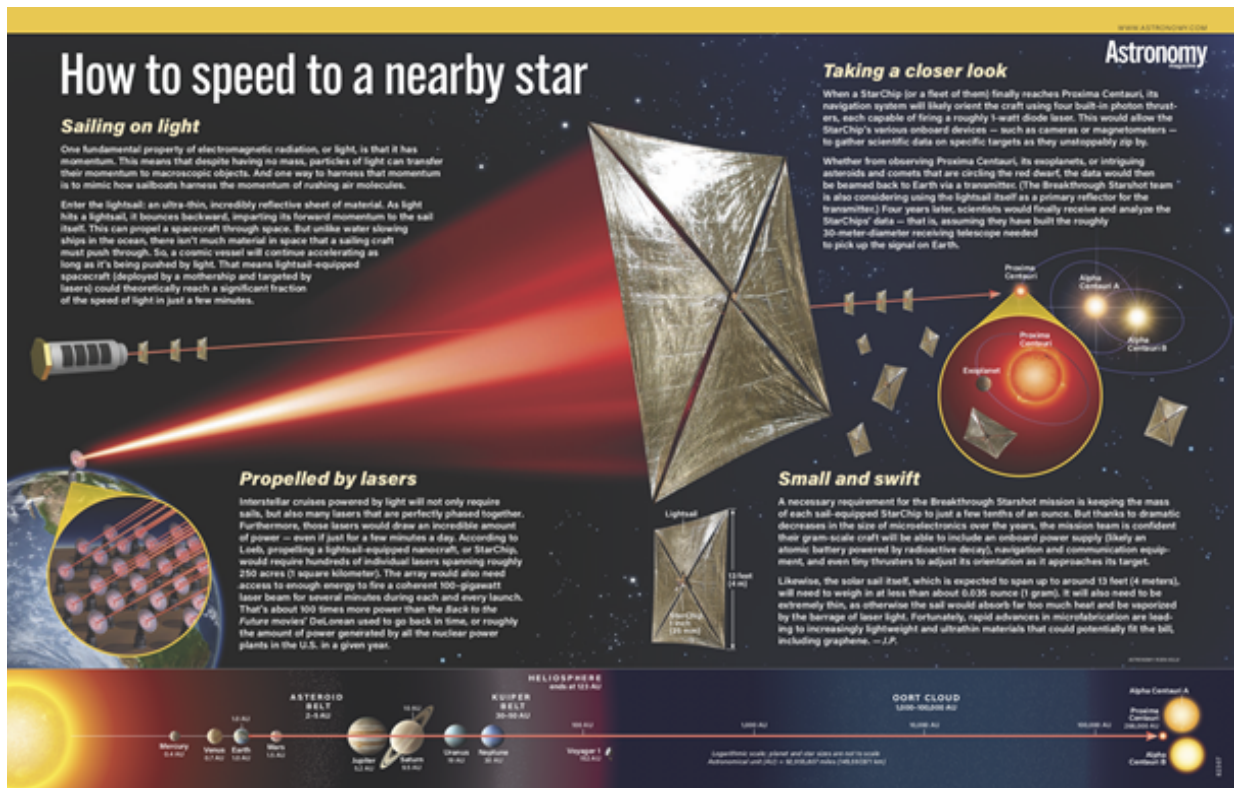
*Figure credit: Astronomy:* Roen Kelly

For the Breakthrough Starshot initiative there are no violations of the known laws of physics that need to occur in order for the mission to succeed. There are a lot of problems to overcome and it is still very challenging. It seems likely to me that it will have to be attempted and much will be learned from the attempt and its failures to make later efforts succeed.

**Problems: Actually make the laser driver system**
All gigawatt-plus lasers in existence only fire for billionths or trillionths of a second. What Starshot is proposing is a 2 minute blast of a 100 gigawatt laser. The closest we have attained to date is 50,000 watts or about a millionth of what is proposed. This challenge is unprecedented; it may not be possible from a materials standpoint as the energy densities just may be too great. A 100 gigawatts for 2 minutes is a staggering number, the output of 100 industrial nuclear reactors simultaneously, but that is only half of the requirement. Energy conversion in laser technology is high around 50%, but to output 100 gigawatts will require 200 gigawatts input power. The energy lost in conversion efficiency turns into excess waste heat a lot of it. A total of 3.33 GWh of waste heat is generated in the 2 minutes that the laser is fired. To put this in perspective, this is the equivalent energy of about a kiloton nuclear explosion.

**Atmospheric distortion**. The military gave up on super high energy lasers due to atmospheric lensing (megawatt size). If the air gets hot enough its starts refracting and reflecting the light. Adaptive optics won't correct this as it is an issue with the atmosphere not being totally transparent and absorbing some of the laser beam as heat. The heat changes the air's index of refraction and bends the beam. The problem gets worse the more power is added to the beam. Dust, moisture, and air turbulence from wind consistently defeat military laser technology. Firing a laser through a thick atmosphere one expects this, as air is not completely transparent. When beaming multi-gigawatt laser beams through air, just a very small fraction of absorption will result in spectacular heating. If the air absorbed only 0.001% of the beams energy that would mean a million watts of heating. This cascades the problem by creating a plasma out of the air which then diffuses the beam rapidly which then leads to more absorption of the beam by the atmosphere. For this to work the lasers would need to be space based.

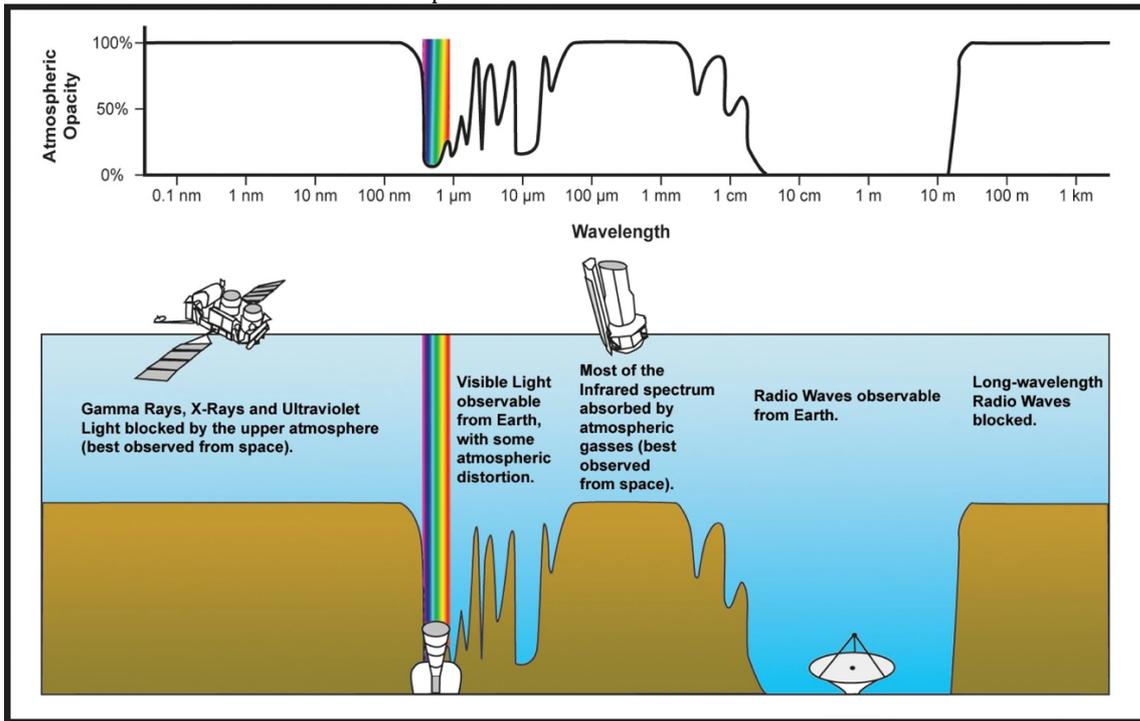

Figure: Atmospheric opacity as a function of wavelength.

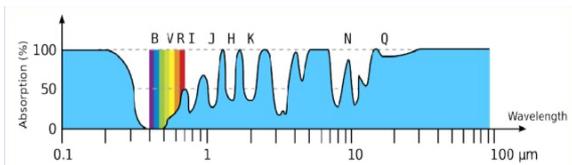

Figure Atmospheric Transmission and choice of astronomical filters

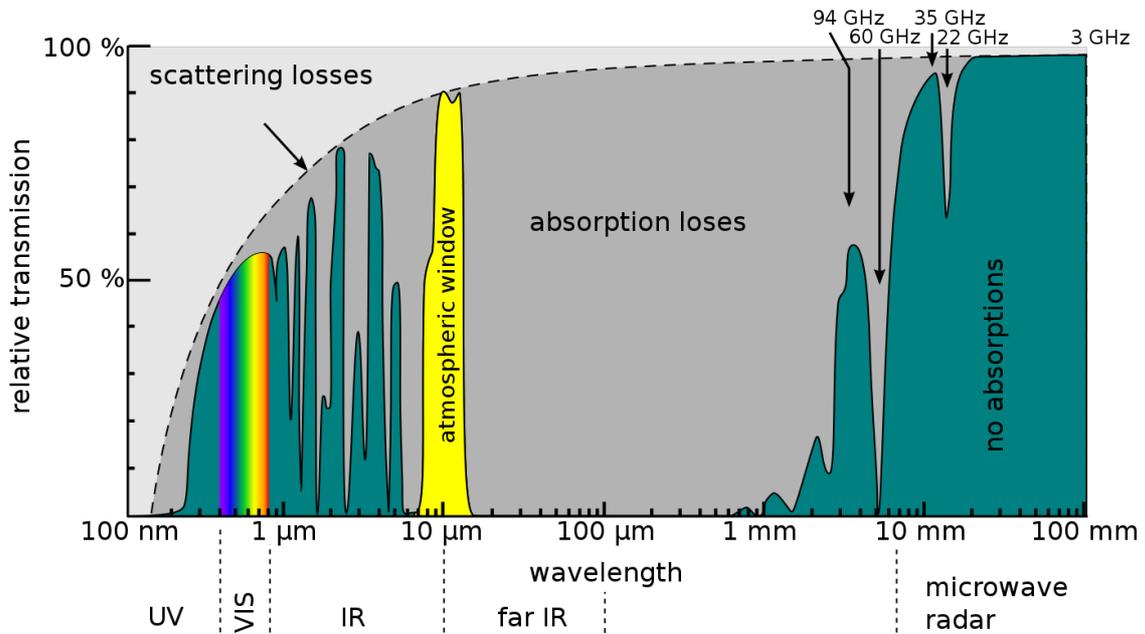

Figure Atmospheric Transmission losses Wikipedia commons. Absorption spectrum during atmospheric transition of electromagnetic radiation. An atmospheric transmission 'window' can be seen between 8-14 μm (700 - 1250 cm$^{-1}$).

The optical window, and thus optical astronomy and laser propulsion, can be severely limited by atmospheric conditions such as clouds and air pollution.

**Problems: Tearing and Melting**
Earlier versions of light sails relied on sunlight alone to power them. But Starshot's first new design would use ground-based lasers to help push the sail to faster speeds. The light intensity would be *millions* of times greater than using sunlight.

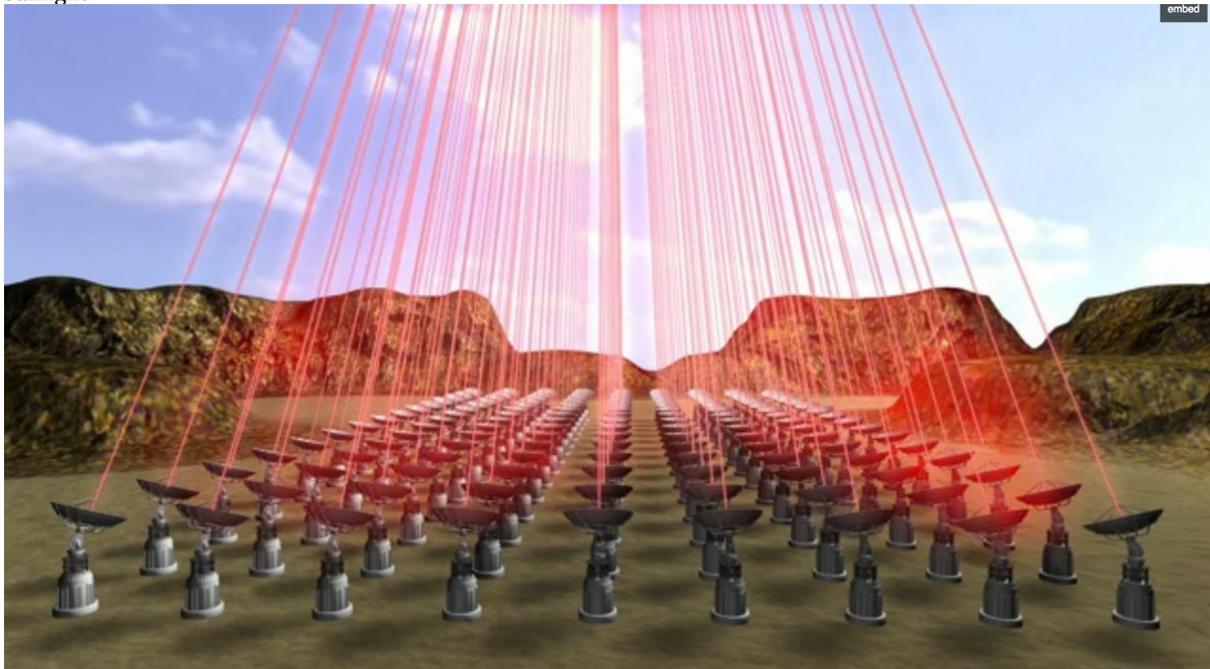

Figure: Concept of ground-based array of lasers to push on light sail. This is an artists drawing as the angles are greatly exaggerated and each beam is likely as wide as the parabolic dish sending it outward. One also has to make real time adjustments to compensate for atmospheric variations. It is the atmospheric variations that make space-based laser system preferable along with the fact that the alignment to send the light sail nano-spacecraft to the desired target(s) will happen much more often.

The general result is that the spacecraft has an Area density: $\sim< 1$ gram/$m^2$ .

Even with our current best adaptive optics and phased array technologies implemented, a terrestrial laser array, even at high altitudes, would need to see improvements of a factor between 10 and 100 to be viable. In addition, even the most reflective surfaces known to humanity — which reflect 99.999% of the energy incident upon them — would currently absorb about ~0.001% of the total energy impacting them. This is, at least at present, is doubly catastrophic.
1. It would incinerate the light sail in short order, rendering it useless and incapable of accelerating to anywhere near the design parameters.
2. The light sail itself, while being accelerated by the incident laser light, would experience a differential force on it across its surface, creating a torque and causing the sail to rotate, making a continuous, directed acceleration impossible. That is, it would be difficult to aim in the correct direction.

**Acceleration**. The record for electronics surviving millisecond high load g forces is 60,000 g's. Now surviving that for 2 minutes is another story. This poses fundamental material problems that may not be overcome. The acceleration would have to come down. Such an acceleration will change the weight of the microchip from a gram to 300kg or 660lbs. How will the sail material and configuration hold a nearly 300 kg load stable while accelerating? It seems a bit of a stretch to the extreme. The likely scenario is that the sail material will be torn to shreds between the g forces and the temperature rise.

To achieve this, the sail must be able to withstand the danger of tearing or melting. To achieve a speed of 0.2c in of order an hour requires an acceleration of a = 0.2c/hour = 60 000 km/s / 60x60/sec =1/60 million m/$s^2$ ~ 1700 gees. If it is to happen in 2 minutes this goes up to a = 0.2c/2 min = 3 x $10^7$ /(60 x 10) g =50 000 gees. So even if the sail's mass is only one gram, the force would be 50 kg (110 lbs). This is very much for a thin sheet (even stretching it).
One solution is to have the sail billow out like a parachute instead of just remaining flat. It would need to be about as deep as it is wide. Such a sail could withstand the strain of hyper-acceleration, a pull thousands of times that of Earth's gravity. Laser photons will fill the sail much like air inflates a beach ball. We also know that lightweight, pressurized containers should be spherical or cylindrical to avoid tears and cracks.
There will be a serious issue with any bumps, wrinkles or ripples in the sail, either static or dynamic, and these would not only threaten to tear it apart but also to ruin the aim and direction of motion.

The laser power illuminating and reflecting from the light sail would need to be
400 kg/4 $m^2$ ~ 100 x $10^3$ g/$m^2$ = Pressure = Energy flux / c so that Energy flux (power per unit area) =
Energy needed for 4 $m^2$ is 400 x $10^3$ gm x 3x$10^8$ m/s = 12 x $10^{10}$ kgm-m/s or 120 gigawatts focused on the sail.

The most reflective material currently available to make the light sail is of order 99.9% reflective. That means that the light sail absorbs power at the rate of order 30 MW (megawatts) or greater. That would heat the mere gram of material at a rate of at least 7 million degrees per second. This quickly gets to melting or vaporizing the light sail unless it can get rid of the energy equally quickly. As there is no conduction, radiation is the only reasonable approach. With unit emissivity i.e. perfect on the back side, the temperature is set by the blackbody radiation of the back. By subsequently integrating over the solid angle for all azimuthal angle (0 to ) and polar angle from 0 to $\pi/2$, we have the Stefan–Boltzmann law: the power $j$ emitted per unit area of the surface of a black body is directly proportional to the fourth power of its absolute temperature: j

Inverting this we have T = (30 x $10^6$ W /= (0.529 x $10^{20}$)K=85 283 K.

This is so high that one has to do the calculations more precisely and to lower the ambient power as well as keep the reflecting surface as perfect as possible.

We have to figure out how to approach the very best case 99.999% reflectivity which would still heat at the rate of $10^{-5}$ x 30GW = 30 000 watts or a rough heating rate of the sail of 30 000 C/second. If we can make the sail have a back side emissivity approaching unity then that means its temperature rises only to T = (30 000 W /= (5.29 x $10^{16}$)K=15 160 K.

**How to dissipate the heat from the laser to keep it from melting.** The answer, the researchers say, is nanoscale patterning (nanolithography). Nanolithography is a branch of nanotechnology and the name of the process for imprinting, writing or etching patterns in a microscopic level in order to create incredibly small structures. The very same process is used to produce CPU chips. These are to maximize their ability to radiate their heat away, which is the only mode of heat transfer available in space.

This is difficult to manage and requires very good quality control.

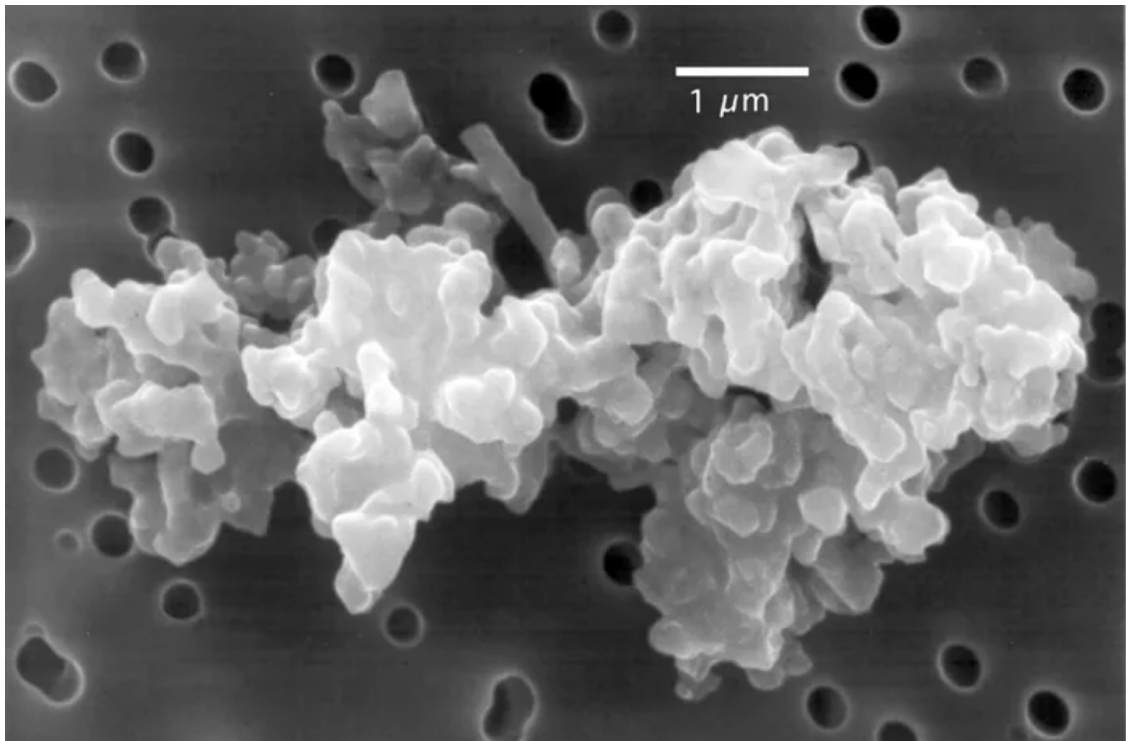

Figure *This scanning electron microscope image shows an interplanetary dust particle at the slightly greater than ~1 micron scale. In interstellar space, we have only inferences about what the dust distribution is, in terms of both size and composition, especially at the low mass and small size end of the spectrum. (Credit: E.K. Jessberger et al., in Interplanetary Dust, 2001)*

**Problems: Destroyed by Dust Collisions**

If Breakthrough Starshot were to go at velocity=0.2c from Earth to the Alpha Centauri system, how many particles (protons, dust grains, etc.) and temperatures would be encountered, and what would be the consequences of each on a thin light sail? And on the working part of the nanocraft?

If the light sail craft is traveling at 0.2c, then a collision with a dust gain may well punch through it or make a major crater in the craft. It could affect not only its orientation but several collisions could alter its course.

The environment and dust of the solar system is fairly well known. However, once the craft leaves the Solar System, the density and size distribution of particles that a traveling spacecraft will encounter changes. The best data we have for that comes from a combination of modeling, remote observations, and direct sampling courtesy of the Ulysses mission. The average density of a cosmic dust particle is about 2.0 grams per cubic centimeter, or about twice the density of water. Most of the cosmic dust particles are tiny and low in mass, but some are larger and more massive.

If you were able to reduce the cross-sectional size of your entire spacecraft to one square centimeter, you'd expect, on a ~4 light-year journey, not to encounter any particles that are ~1 micron or larger in diameter. You would only have about a 10% chance of doing so. However, as you look to smaller particles, you start to anticipate a much larger number of collisions: 1000 collisions with particles about ~0.1 microns in diameter, 10,000 collisions with particles about ~0.05 microns in diameter, 100,000 collisions with particles about ~0.03 microns in diameter, 1,000,000 collisions with particles about ~0.018 microns in diameter, 10,000,000 collisions with particles about ~0.01 microns in diameter. You might think this is not a big deal, to encounter such a large number of such tiny particles, especially when you consider how minuscule the mass of such particles would be. For example, the largest particle you'd hit, at 0.5 microns in diameter, would only have a mass of about 4 picograms ($4 \times 10^{-12}$ g). By the time you got down to a particle of ~0.1 microns in diameter, its mass would be a paltry 20 femtograms ($2 \times 10^{-14}$ g). And at a size of ~0.01 microns in diameter, a particle would only have a mass of 20 attograms ($2 \times 10^{-17}$ g).

But this, when you do the math, is disastrous. It isn't the biggest particles that impart the most energy to a spacecraft traveling through the interstellar medium, but the smallest ones. At 20% the speed of light, a ~0.5 micron diameter particle will impart 7.2 Joules of energy to this tiny spacecraft, or about as much as energy as it takes to raise a 5 pound (~2.3 kg) weight from the ground to over your head.

Now, a ~0.01 micron diameter particle, also moving at ~20% the speed of light, will only impart 36 micro-Joules of energy to that same spacecraft: what seems like a negligible amount.

But these latter collisions are *10 million times* more frequent than the largest collisions expected to occur. When we look at the total energy loss anticipated from dust grains that are ~0.01 microns or larger, it's straightforward to calculate that there are a total of about ~800 Joules of energy that will be deposited into every square centimeter of this spacecraft from collisions with the various-sized dust particles in the interstellar medium.

Even though it will be spread out, in time and over the cross-sectional area of this tiny spacecraft, that's a tremendous amount of energy for something that has a mass of only ~1 gram or so. It teaches us a few valuable lessons:

1. The current Breakthrough Starshot idea, of applying a protective coating of a material like beryllium copper to the nanocraft, is wildly insufficient.
2. The laser sail will be in danger of becoming absolutely shredded in short order and will also cause a substantial drag on the nanocraft if it isn't jettisoned or (somehow) folded and stowed after the initial laser-driven acceleration takes place.
3. Collisions from even smaller objects—things like the molecules, atoms, and ions that exist throughout the interstellar medium — will add up as well, and will potentially have even greater cumulative effects than dust particles will.

There are, of course, potentially clever solutions to many of these problems that are available. For example, if you determined that the light sail itself would suffer too much damage or would slow down your journey by too great an amount, you could simply detach it once the laser acceleration stage was complete. If you designed your nanocraft — the "spacecraft" part of the apparatus — to be very thin, you could direct it to travel so that its cross-section in the direction of motion was minimized. And if you determined that the damage from ions would be substantial, you could potentially set up a continuous electric current through the spacecraft, generating its own magnetic field to deflect charged cosmic particles. This would be power probative unless one uses superconductors and rapidly is an issue.

800 joules energy imparted also slows the one gram craft ½ m$\delta v^2$=800 or $\delta v$ = (2x800joule/0.001kgm)$^{1/2}$ ~ 1300 m/s ~ 0.004c. There will be some scatter in the number of particles which collide giving a small spread to this speed at the hundreth of percent level. In a trip to Proxima b that difference in speed will slow and decrease the distance travelledby the average 1 cm² craft by about 0.04 light years or 9.4 x 0.04x10$^{12}$ km = 0.4 x 10$^{12}$ km and thus spread out the location along the travel direction of the cluster of craft by roughly 40 million km. This is a few times the 7.5 million km orbital radius of Proxima b around Proxima Centauri.

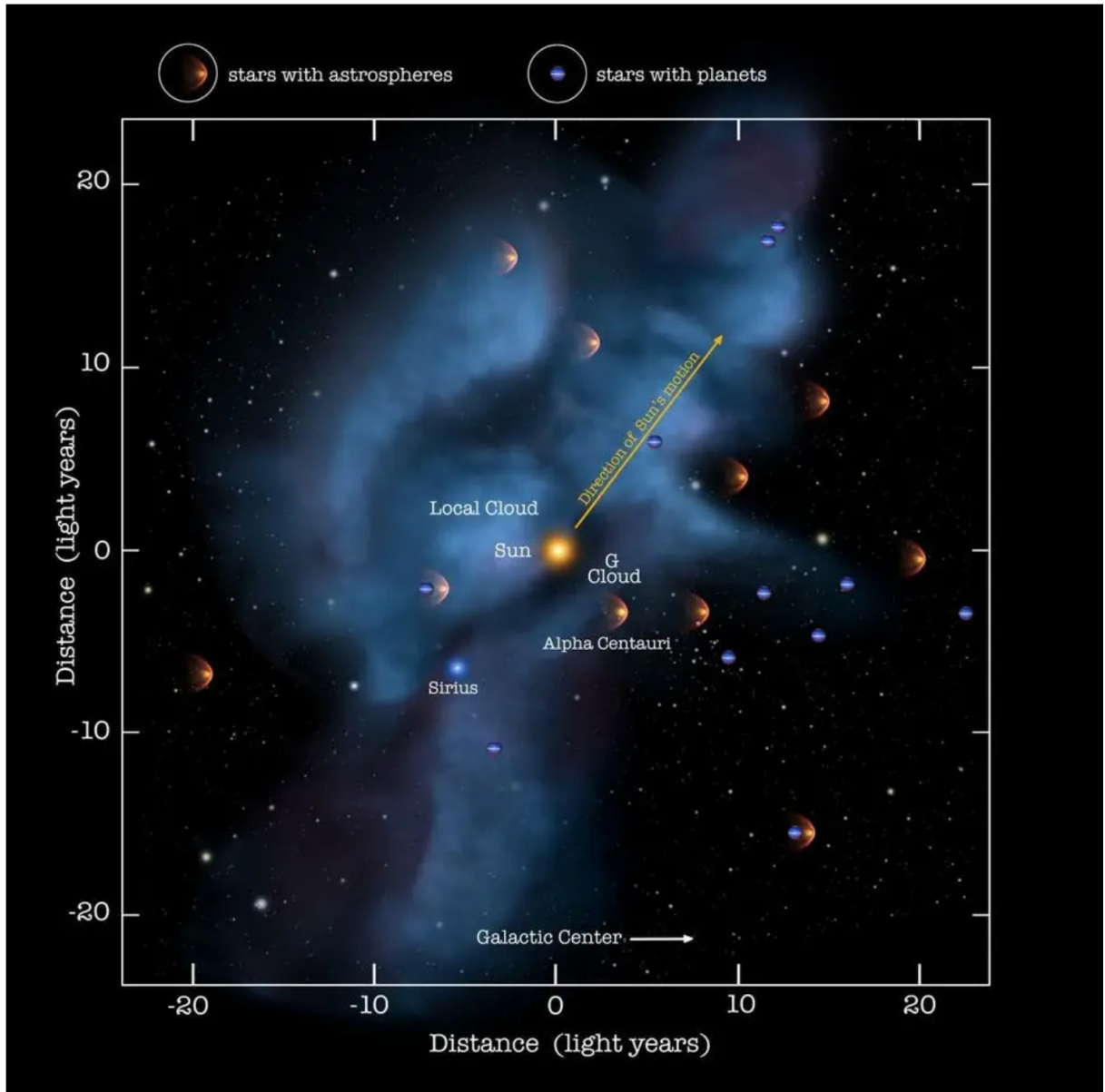

*The copious amount of space between the stars and star systems in our neighbourhood isn't completely empty, but is filled with gas, dust, molecules, atoms, ions, photons, and cosmic rays. The faster we move through it, the more damage we'll incur, irrespective of the size or composition of our spacecraft. (Credit: NASA/Goddard/Adler/U. Chicago/Wesleyan)[6]*

**Multiple NanoCraft**
Once a laser system is created and the first light nano-spacecraft is manufactured along with a way to deploy into space, then it does not cost much extra to make very many nano-spacecraft and fire the laser array. It is not free but is very cheap to make many compared to the initial costs of making the first. So, people have considered launching several spacecraft towards each of our 50 nearest neighbourhood extra-solar systems or the current baseline of sending 1000 nano-craft towards Proxima-b. Drop them out and laser boost one after the other. One can both stagger the nano-craft launch but boost with slightly increasing power so that they all arrive relatively together at Proxima-b.

They will still be a bit of spread around the Proxima system as it is hard to aim that accurately or to correct their course that accurately. However one is likely to have coverage of the system and having multiple craft makes it more likely that some survive the hazards of the journey such as hitting dust particles. There may be random perturbations to the trajectory such that all the paths become randomized during the travel to Proxima b, thereby setting a lower limit to the number of craft required to guarantee coverage of the system)

However, once one has an array of craft and sensors as well as distributed intelligence, it is possible to survey the stellar planetary system more thoroughly and perhaps to communicate back to Earth area more effectively.[7]

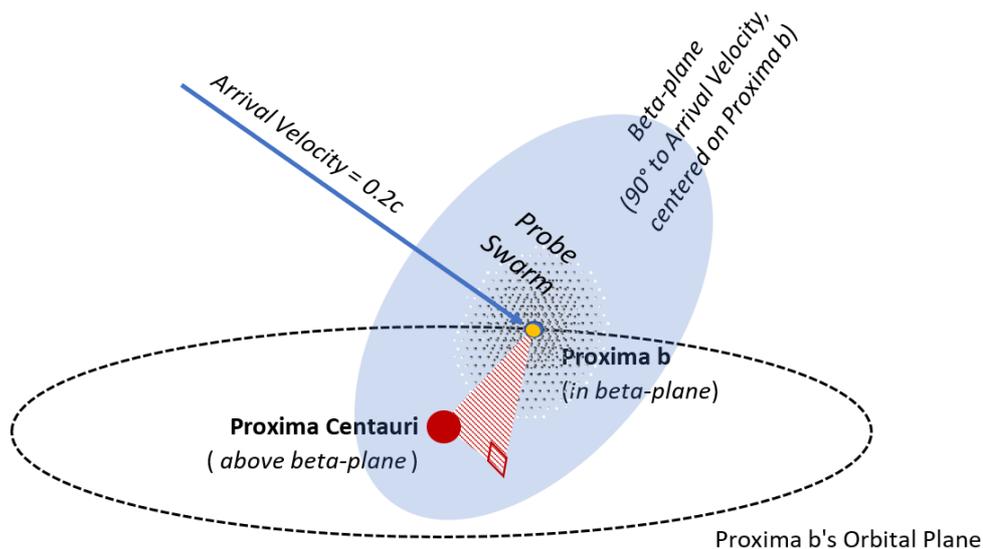

Figure: Geometry of swarm's encounter with Proxima b. The Beta-plane is the plane orthogonal to the velocity vector of the probe "at infinity" as it approaches the planet; in this example the star is above (before) the Beta-plane. To ensure that the elements of the swarm pass near the target, the probe-swarm is a disk oriented perpendicular to the velocity vector and extended enough to cover the expected transverse uncertainty in the probe-Proxima b ephemeris.

**Supercapacitors**

Although supercapacitors present lower energy densities than batteries, they possess numerous assets that batteries do not; primarily a virtually infinite lifetime. They typically store 10 to 100 times more energy per unit volume or mass than electrolytic capacitors, can accept and deliver charge much faster than batteries, and tolerate many more charge and discharge cycles than rechargeable batteries.

Supercapacitors are more efficient than batteries, especially under full load conditions. They can achieve round-trip efficiency of more than 98% while lithium-ion batteries' efficiencies are less than 90%. For example, graphene has a theoretical specific surface area of 2630 $m^2$/g which can theoretically lead to a capacitance of 550 F/g. Supercapacitors can provide a big burst of power for data transmission or other tasks.

Electrochemical capacitors (supercapacitors) consist of two electrodes separated by an ion-permeable membrane (separator), and an electrolyte ionically connecting both electrodes. When the electrodes are polarized by an applied voltage, ions in the electrolyte form electric double layers of opposite polarity to the electrode's polarity. For example, positively polarized electrodes will have a layer of negative ions at the electrode/electrolyte interface along with a charge-balancing layer of positive ions adsorbing onto the negative layer. The opposite is true for the negatively polarized electrode.

Additionally, depending on electrode material and surface shape, some ions may permeate the double layer becoming specifically adsorbed ions and contribute with pseudocapacitance to the total capacitance of the supercapacitor.

| Development | Date | Specific energy[A] | Specific power | Cycles | Capacitance | Notes |
|---|---|---|---|---|---|---|
| Laser-induced graphene/solid-state electrolyte[1,2] | 2015 | | 0.02 mA/$cm^2$ | | 9 mF/$cm^2$ | Survives repeated flexing. |

1. "Flexible 3D graphene supercapacitors may power portables and wearables | KurzweilAI". www.kurzweilai.net. Retrieved 11 February 2016.
2. Peng, Zhiwei; Lin, Jian; Ye, Ruquan; Samuel, Errol L. G.; Tour, James M. (28 January 2015). "Flexible and Stackable Laser-Induced Graphene Supercapacitors". ACS Applied Materials & Interfaces. **7** (5): 3414–3419. doi:10.1021/am509065d. PMID 25584857.

By simple laser induction, commercial polyimide films can be readily transformed into porous graphene for the fabrication of flexible, solid-state supercapacitors. Two different solid-state electrolyte supercapacitors are described, namely vertically stacked graphene supercapacitors and in-plane graphene microsupercapacitors, each with enhanced electrochemical performance, cyclability, and flexibility. Devices with a solid-state polymeric electrolyte exhibit areal capacitance of >9 mF/$cm^2$ at a current density of 0.02 mA/$cm^2$, more than twice that of conventional aqueous electrolytes. Moreover, laser induction on both sides of polyimide sheets enables the fabrication of vertically stacked supercapacitors to multiply its electrochemical performance while preserving device flexibility.

**Laser Communication System**

Historically spacecraft have used radio systems for communication. There also were arguments in early days that the waterhole, or water hole, is an especially quiet band of the electromagnetic spectrum between 1420 and 1662 megahertz, corresponding to wavelengths of 21 and 18 centimeters, respectively. It is a popular observing frequency used by radio telescopes in radio astronomy. It is very low energy cost per photon as well as minimum background radiation.

The strongest [hydroxyl radical](#) [spectral line](#) radiates at 18 centimeters, and atomic [hydrogen](#) at 21 centimeters (the [hydrogen line](#)). These two molecules, which combine to form [water](#), are widespread in [interstellar gas](#), which means this gas tends to absorb radio noise at these frequencies. Therefore, the spectrum between these frequencies forms a relatively "quiet" channel in the interstellar [radio noise](#) background. The low frequency f and low background means that the number of photons n and energy ($E=nhf=nhc/\lambda$) is minimal to get information transmitted reliably.

Typical semiconductor lasers have wavelengths $\lambda$ on the scale of 500 nm or about $\lambda \sim 0.2$ m/$0.5 \times 10^{-6}$ m = $0.4 \times 10^{6}$ times more energy per photon and more background photons from the alien system star unless the frequency is tailored to fit in a low signal region or made extremely narrow bandwidth. However, in principle one can gain back the large wavelength factor, if the laser beam can be focused as well as the radio.. For a characteristic 1-m size, a laser focus could be down to $\delta\theta \sim 1.22\, \lambda / d \sim 0.5 \times 10^{-6}$ radians. At 4.2 light years this is $4.2 \times 3.15 \times 10^{7} \times 0.5 \times 10^{-6}$ light sec $\sim$ 66 light sec $\sim$ 2 million km or the size of the inner solar system. The issue will be for the AI to get the feedback to keep the laser communication system pointed sufficiently accurately. Testing and additional feedback may make that possible.

Basically if one could focus the optical laser as effectively as radio telemetry then the scaling with wavelength would cancel each other. I.e. if the diameter of the transmitting dish is d (which we take to be on the one meter scale), then the relative beam widths scales as the ratio of the wavelengths – as is set by diffraction $\delta\theta = 1.22\, \lambda/d$. The distance from Proxima b to Earth is L = 4.2 lightyears = $4.2 \times 9.4607 \times 10^{15}$ [m](#). Thus the effective spot diameter $L\, \delta\theta = L\, \lambda/d$ and the effective area is roughly $A = \pi\, (L\, \lambda/d)^2$. If each transmitter has the same efficiency, the number of photons sent at each wavelength would be $N = E\lambda/hc$. The number of photons per unit area is roughly n.So if the detecting receiver has diameter D independent of $\lambda$, then the number of photons detected would maximally be $N/S = E\, \lambda\, /[hc\, \pi\, (L\, \lambda/d)^2\,] = E\, d^2\, /[\, \lambda\, hc\, \pi\, L^2\,]$. We see the number of photons detected increases with frequency or one over the wavelength. After the waterhole the background photons generally increase with the frequency. Thus depending upon the actual engineering and details it may be an advantage to move up to laser frequencies and short wavelengths.

## Moderately changed approach

Initially Yuri Milner's interest in Breakthrough Starshot was to learn much more about possible alien worlds in his lifetime. This desire to have the results in a human lifetime is what drives the goal for a speed of 0.2c which results in many severe demands on the overall system that are fierce challenges.

What if Mr. Milner invests some of his fortune in life expectancy so that he could easily live for another 100 years? Then we could reduce the necessary speed a factor of 5 to 0.04c. This simple change reduces our challenges by factors of v and $v^2$ or factors of 5 to 25. It does require a monitoring and operating institution whose lifetime must increase as 4.2 years x (v/c + 1) which is $10^9$ years for this case of v = 0.04c.

Dropping the needed laser power by a factor of 5 makes the cost and complexity of the laser booster system decrease by more than that factor of 5 bringing it down from at least 0.9 billion$ to the scale of 200M$ and closer in range of existing technology. The reduction in laser power reduces the heating of the light sail by that same factor of 5, decreasing the radiating temperature by a factor of 1.5 and reducing mechanical stress by the same factor of 5.

The largest gain comes from the decreased effect of impacts by dust and molecules which is proportional to the relative kinetic energy and goes as $v^2$ so these are reduced by a factor of 25.
The disadvantage is that it lengthens the exposure to cosmic rays by about a factor of 5 and also component failure due to aging. There is likely a trade off in these factors that will have a minimum at a speed between the $10^{-4}$c (max of rocket launched spacecraft) < v/c < 0.05 and closer to 0.01 than the chemical rocket with gravity assist ends.

# Conclusions

"We believe it will take at least several decades to develop the necessary technology - but would hope to launch our first interstellar probes within 25–40 years." - Simon Pete Worden, Executive Director, Breakthrough Initiatives.

This is likely an appropriate but optimistic assessment of the goal to have a technology set and capability for an interstellar probe that could make the trip to an extrasolar system and send back observations. If one were not extremely optimistic, one would not start such a project. While the chances of such an attempt is likely much further in the future, it is good to consider the mission and develop the concepts and technologies that would be needed. It might do well to consider a mission that would take longer, i.e. be a bit slower, to handle these difficult challenges.

**Acknowledgements:** I thank my colleagues on the interstellar photovoltaics paper and the interest, checking and comments from Amruth Alfred.
No funding for this or related from Breakthrough Starshot.